\documentclass{article}
\usepackage{amsmath}
\usepackage{amssymb}
\usepackage{amsfonts}
\usepackage{amsthm}
\usepackage{graphicx}

\def\dbar{{\mathchar'26\mkern-12mu d}}
\newcommand{\un}{\hat{n}}						
\newcommand{\vp}{\times}	
\newcommand{\dpi}{\pi_{ij}}
\newcommand{\dtau}{\tau_{ij}}
\newcommand{\udyad}{\delta_{ij}}				
\newcommand{\dsigma}{\sigma_{ij}}
\newcommand{\pdt}[2]{\frac{\partial{#1}}{\partial{#2}}}		

\newcommand{\grad}[1]{\vec{\nabla}{#1}}
\newcommand{\dive}[1]{\vec{\nabla}\cdot{#1}}
\newcommand{\curl}[1]{\vec{\nabla}\vp{#1}}

\begin{document}
\title{Stress due to Electric and Magnetic fields in Viscoelastic Fluids}
\author{Amey Joshi\\Tata Consultancy Services,\\International Technology Park,\\Whitefield, Bangalore, 560 066\\}
\maketitle

\begin{abstract}
A clear understanding of body force densities due to external electromagnetic fields is necessary to study flow and deformation of materials exposed to the fields. In this paper, we 
derive an expression for stress in continua with viscous and elastic properties in presence of external, static electric or magnetic fluids. Our derivation follows from fundamental
thermodynamic principles. We demonstrate the soundness of our results by showing that they reduce to known expressions for Newtonian fluids and elastic solids. We point out the extra
care to be taken while applying these techniques to permanently polarized or magnetized materials and derive an expression for stress in a ferro-fluid. Lastly, we derive expressions for
ponderomotive forces in several situations of interest to fluid dynamics and rheology.
\end{abstract}

\section{Introduction}\label{sec:intro}
We can study the effect of electromagnetic fields on fluids only if we know the stress induced due to the fields in the fluids. Despite its importance, this topic is glossed over in most 
works on the otherwise well-established subjects of fluid mechanics and classical electrodynamics. The resultant force and torque acting on the body as a whole are calculated but not the 
density of body force which affects flow and deformation of materials. Helmholtz and Korteweg first calculated the body force density in a Newtonian dielectric fluid in the presence of an
electric field, in the late nineteenth century. However, their analysis was criticized by Larmor, Livens, Einstein and Laub, who favoured a different expression proposed by Lord Kelvin. It 
was later on shown that the two formulations are not contradictory when used to calculate the force on the body as whole and that they can be viewed as equivalent if we interpret the 
pressure terms appropriately. We refer to Bobbio's treatise~\cite{ch2:bobbio} for a detailed account of the controversy, the experimental tests of the formulas and their eventual 
reconciliation. The few published works on the topic like the text books of Landau and Lifshitz~\cite{ch2:ll8}, Panofsky and Phillips~\cite{ch2:pp} and even Bobbio~\cite{ch2:bobbio} treat 
fluids and elastic solids separately. Further, they restrict themselves to electrically and magnetically linear materials alone. In this paper, we develop an expression for stress due to 
external electromagnetic fields for materials with simultaneous fluid and elastic properties and which may have non-linear electric or magnetic properties. Our analysis is thus able to 
cater to dielectric viscoelastic fluids and ferro-fluids as well. We also extend Rosensweig's treatment~\cite{ch2:rer}, by allowing ferro-fluids to have elastic properties.

Let us first see why the problem of finding stress due to electric or magnetic fields inside materials is a subtle one while that of calculating forces on torques on the body as a 
whole is so straightforward. The standard approach in generalizing a collection of discrete charges $q_i$ to a continuous charge distribution is to replace the charges themselves with a 
suitable density function $\rho_e$ and sums by integrals. Thus, the expression for force $\sum_{i=1}^Nq_i\vec{E}_i$, ($\vec{E}_i$ is the electric field at the location of the charge 
$q_i$.) on a body on $N$ discrete charges in an electric field $\vec{E}$, is replaced with $\smallint\rho\vec{E}(\vec{x})dV$, when the body is treated as a continuum of charge, the 
integral being over the volume of the body. The integral can be written as
\begin{equation}\label{intro:1}
\int_V\rho\vec{E}(\vec{x})dV = \int_V\vec{f}dV
\end{equation}
where $\vec{f}=\rho_e\vec{E}$ is the force density in the body due to an external electric field. It can be shown that~\cite{ch2:bobbio} that the same expression for force
density is valid even inside the body. If instead, the body were made up of discrete dipoles instead of free charges, then the force on the body as a whole would be written 
as~\cite{ch2:rmc}
\begin{equation}\label{intro:2}
\vec{F}=\sum_{i=1}^N \vec{p}_i\cdot\grad{\vec{E}_i},
\end{equation}
where $\vec{p}_i$ is the dipole moment of the $i$th point dipole and $\vec{E}_i$ is the electric field at its position. If the body is now approximated as a continuous distribution of 
dipoles with polarization $\vec{P}$, then the force on the whole body is written as
\begin{equation}\label{intro:3}
\vec{F}=\int\vec{P}\cdot\grad{\vec{E}}dV,
\end{equation}
While this is a correct expression for force on the body as a whole, it is not valid if applied to a volume element inside the material. In other words, $\vec{P}\cdot\grad{\vec{E}}$ is 
not a correct expression for density of force in a continuous distribution of dipoles although $\rho_e\vec{E}$ is the density of force in the analogous situation for monopoles. We shall 
now examine why it is so.

Consider two bodies $\mathcal{B}_c$ and $\mathcal{B}_d$ that are composed of charges and dipoles respectively. (The subscripts of quantities indicate their composition.) Let $V_c$ and 
$V_d$ be volume elements of $\mathcal{B}_c$ and $\mathcal{B}_d$ respectively. The volume elements are small compared to dimensions of the body but big enough to have a large number of 
charges or dipoles in them. The forces $\vec{F}_c$ and $\vec{F}_d$ on $V_c$ and $V_d$ respectively due to the surrounding body are
\begin{eqnarray}
\vec{F}_c &=& \sum_{i=1}^Nq_i\vec{E}_i	\\
\vec{F}_d &=& \sum_{i=1}^N\vec{p}_i\cdot\grad{\vec{E}_i},
\end{eqnarray}
where $N$ is the number of charges or dipoles inside the volume element under consideration.
In both these expressions, $\vec{E}_i$ is the macroscopic electric field at the position of $i$th charge or dipole. It is the average value of the microscopic electric field 
$\vec{e}_i$ at that location. That is $\vec{E}_i=\langle\vec{e}_i\rangle$, where $\langle\cdot\rangle$ denotes the spatial average of the enclosed quantity. The microscopic field 
$\vec{e}_i$ can be written as $\vec{e}_i^{\text{ }ext}+\vec{e}_i^{\text{ }int}$ where $\vec{e}_i^{\text{ }ext}$ is the microscopic field due to the charges or dipole outside the volume 
element and $\vec{e}_i^{\text{ }int}$ is the field due to charges or dipoles inside the volume element other than the $i$th charge or dipole. For the volume element $V_c$ of point charges,
\begin{equation}
\vec{e}_i^{\text{ }int} = \sum_{j=1,j \ne i}^N\vec{e}_{i,j},
\end{equation}
where $\vec{e}_{i,j}$ is the microscopic electric field at the position of $i$th charge due to $j$th charge inside $V_c$. Therefore,
\begin{equation}
\vec{F}_c = \sum_{i=1}^Nq_i\langle\vec{e}_i^{\text{ }ext}\rangle + \left\langle\sum_{i=1}^Nq_i\sum_{j=1,j \ne i}\vec{e}_{i,j}\right\rangle
\end{equation}
Newton's third law makes the second sum on the right hand side of the above equation zero. $\vec{F}_c$ is thus due to charges outside $V_c$ alone for which the standard approach of 
replacing sum by integral and discrete charge by charge density is valid. Therefore, $\rho_e\vec{E}$ continues to be the volume force density inside the body. If the same
analysis were to be done for the volume element $V_d$ of point dipoles, it can be shown that the contribution of dipoles inside $V_d$ is not zero. In fact, the contribution depends on the
shape of $V_d$~\cite{ch2:jdj}. That is the reason why $\vec{P}\cdot\grad{\vec{E}}$, also called Kelvin's formula, is not a valid form for force density in a dielectric material. 

We would have got the same results for a continuous distribution of magnetic monopoles, if they had existed, and magnetic dipoles. That is $\vec{M}\cdot\grad{\vec{H}}$ is not the correct
form of force density of a volume element in a material with magnetization $\vec{M}$ in a magnetic field $\vec{H}$. The goal of this paper is to develop an expression for stress 
inside a material with both viscous and elastic properties in the presence of an external electric or magnetic field, allowing the materials to have non-linear electric and magnetic 
properties. 

We demonstrate that by making some fairly general assumptions about thermodynamic potentials, it is possible to develop a theory of stresses for materials with fluid and 
elastic properties. We check the correctness of our results by showing that they reduce to the expressions developed in earlier works when the material is a classical fluid or solid. To
our knowledge, there is no theory of electromagnetic stresses in general continua with simultaneous fluid and elastic properties. 

Since we are using techniques of equilibrium thermodynamics for our analysis, we will not be able to get results related to dissipative phenomena like viscosity. Deriving an expression for 
viscosity for even a simple case of a gas requires full machinery of kinetic theory~\cite{ch2:cc}. Developing a theory of electro and magneto viscous effects is a much harder problem and 
we shall not attempt to solve it in this paper.

We begin our analysis in section (\ref{sec:thermod}) by reviewing expressions for the thermodynamic free energy of continua in electric and magnetic fields. After pointing out the 
relation between stress and free energy in section (\ref{sec:dielectric}), we obtain a general relation for stress in a dielectric material in presence of an electric field. We check 
its correctness by showing that it reduces to known expressions for stress in Newtonian fluids and elastic solids. The framework for deriving electric stress is useful for deriving 
magnetic stress in materials that are not permanently magnetized. Section (\ref{sec:magnetic}) mentions the expression for stress in a continuum in presence of a static magnetic field. 
We then point out the assumptions in derivations of (\ref{sec:dielectric}) and (\ref{sec:magnetic}) that render the expressions of stress unsuitable for ferro-fluids and propose the one 
that takes into account the permanent magnetization of ferro particles. We derive expressions for ponderomotive forces in section (\ref{sec:ponder}) from the expressions for stress 
obtained in previous sections. Most of our analysis rests on framework scattered in the classic works of Landau and Lifshitz on electrodynamics~\cite{ch2:ll8} and elasticity~\cite{ch2:ll7} 
generalizing it for continua of arbitrary nature. 

\section{Thermodynamics of continua in electromagnetic fields}\label{sec:thermod}
Electromagnetic fields alter thermodynamics of materials only if they are able to penetrate in their bulk. Conducting materials have plenty of free charges to shield their interiors from
external static electric fields. Therefore, the effect of external static electric fields are restricted to their surface alone, in the form of surface stresses. The situation in 
dielectrics is different - a paucity of free charges allows an external static field to penetrate throughout its interior polarizing its molecules. The external field has to do work to 
polarize a dielectric. This is akin to work done by an external agency in deforming a body. The same argument applies to a body exposed to a magnetic field. Unlike static electric fields 
that are shielded in conductors, magnetic fields always penetrate in bodies, magnetizing them. The nature of the response depends on whether a body is diamagnetic, paramagnetic or 
ferromagnetic. In all the cases, magnetic fields have to do work to magnetize them and therefore the thermodynamics of continua is always affected by a magnetic field. We shall develop
thermodynamic relations for materials exposed to static electromagnetic fields in this section.

At a molecular level, electric and magnetic fields deform matter for which the fields have to do work. The material and the field together form a thermodynamic system. The work done on it 
is of the form $\dbar W = X\dbar Y$ where $X$ is an intensive quantity and $Y$ a related extensive quantity\footnote{$\dbar X$ denotes the possibly inexact differential of a quantity $X$.}. In the case of a dielectric material in a static electric field, the intensive quantity is the electric field $\vec{E}$ and the extensive quantity is the total dipole moment $(\vec{P}V)$, $\vec{P}$ being the polarization and $V$ being the volume of the material. In the case of a material getting magnetized, the intensive quantity is the magnetizing field $\vec{H}$ and the extensive quantity is the total magnetic moment $(\vec{M}V)$, $\vec{M}$ being the polarization and $V$ being the volume of the material. The corresponding work amounts are $d\mathcal{W}_1 = \vec{E}\cdot d(\vec{P}V)$ and $d\mathcal{W}_1 = \vec{H}\cdot d(\vec{M}V)$ respectively. We added a subscript '$1$' because this is only one portion of 
the work. The other portion of the work is required to increment the fields themselves to achieve a change in polarization or magnetization. They are $(\epsilon_0\vec{E}\cdot d\vec{E})V$ and $(\mu_0\vec{H}\cdot d\vec{H})V$ respectively, where $\epsilon_0$ is the permittivity of free space and $\mu_0$ is the permeability of free space respectively\footnote{The energy density of an electrostatic field is $u_E = (1/2)\epsilon_0 E^2$ and that of a magnetic field is $u_B = (1/2)\mu_0 H^2$.} Therefore, the total work needed to polarize and magnetize a material, at constant volume, are
\begin{eqnarray}
\dbar \mathcal{W}_e =& \vec{E}\cdot d(\vec{P}V) + (\epsilon_0\vec{E}\cdot d\vec{E})V &= (\vec{E}\cdot d\vec{D}) V	\label{thermod:e1}		\\
\dbar \mathcal{W}_m =& \vec{H}\cdot d(\vec{M}V) + (\mu_0\vec{H}\cdot d\vec{H})V &= (\vec{H}\cdot d\vec{B}) V		\label{thermod:e2}	
\end{eqnarray}
We have derived these relations for linear materials. We will now show that they are true for any material. 

\subsection{Work done during polarization}\label{sec:work_p}
Imagine a dielectric immersed in an electric field. Let the electric field be because of a charge density $\rho_e(\vec{x})$. Let the electric field be increased slightly by changing the 
charge density by an amount $\delta\rho_e$. Work done to accomplish this change is
\begin{equation}\label{thermod:e3}
\delta \mathcal{W}_e = \int \delta\rho_e(\vec{x})\Phi(\vec{x})dV,
\end{equation}
where $\Phi(\vec{x})$ is the electric potential. Since $\dive{\vec{D}} = \rho_e$,
\begin{equation}\label{thermod:e3a}
\delta \mathcal{W}_e = \int (\dive{\delta\vec{D}})\Phi(\vec{x})dV = \int\dive{(\Phi\delta\vec{D})}dV + \int\delta\vec{D}\cdot\vec{E}dV
\end{equation}
If the charge density is localized then the volume of integration can be taken as large as we like. We do so and also convert the first integral on the right hand side to a surface
integral. The first term then makes a vanishingly small contribution to the total and the work done in polarizing a material can be written as
\begin{equation}\label{thermod:e3b}
\delta \mathcal{W}_e = \int\delta\vec{D}\cdot\vec{E}dV
\end{equation}

\subsection{Work done during magnetization}\label{sec:work_m}
Let a material be magnetized by immersing it in a magnetic field. The magnetic field can be assumed to be created because of a current density $\vec{J}$. Let the magnetic field be
increased slightly by changing the current density. We further assume that the rate of increase of current is so small that $\dive{\vec{J}}=0$ at all stages. The source of current has to 
do an additional work while increasing the amount of current density in order to overcome the opposition of the induced electromotive force. If $\delta V$ is the induced emf, then the 
sources will have to do an additional work at the rate $\dot{\mathcal{W}}_m = -I\delta V = I\dot{\Phi}_B$, where $\Phi_B$ is the magnetic flux and the dot over head denotes total time
derivative. The amount of work needed is 
$\delta\mathcal{W}_m = I\Phi_B$. If $\delta s$ is the cross sectional area of the current, then
\begin{equation}\label{thermod:e4}
\delta \mathcal{W}_m = \vec{J}\delta s\int \delta{\vec{B}}\cdot\un dA = \vec{J}\delta s\int(\curl{\delta\vec{A}})\cdot\un dA = \vec{J}\delta s\oint\delta\vec{A}\cdot d\vec{x}
\end{equation}
But $\vec{J}\delta s\cdot d\vec{x} = \vec{J}dV$, therefore,
\begin{equation}\label{thermod:e5}
\delta \mathcal{W}_m = \int\delta\vec{A}\cdot\vec{J} dV
\end{equation}
Since we assumed the current to be increased at an infinitesimally slow rate, there are no displacement currents and $\vec{J}=\curl{\vec{H}}$.
\begin{equation}\label{thermod:e6}
\delta \mathcal{W}_m = \int\delta\vec{A}\cdot(\curl{\vec{H}}) dV = \int\vec{H}\cdot(\curl{\delta\vec{A}})dV - \int\dive{(\vec{H}\vp\delta\vec{A})}dV,
\end{equation}
where we have used the vector identity $\dive{(\vec{F}\vp\vec{G})}=\vec{G}\cdot\curl{\vec{F}}-\vec{F}\cdot\curl{\vec{G}}$. We once again assume that the current density is localized and
therefore converting the second integral on the right hand side of equation (\ref{thermod:e6}) into a surface integral results in an infinitesimally small quantity. The work done in
magnetizing a material is therefore,
\begin{equation}\label{thermod:e7}
\delta \mathcal{W}_m = \int\vec{H}\cdot\delta\vec{B}dV 
\end{equation}

\subsection{Free energy of polarized and magnetized media}\label{sec:fe}
A change in the Helmholtz free energy of a system is equal to the work done by the system in an isothermal process, which in turn is related to stresses in the continuum. We will show how
stress is related to the Helmholtz free energy. Let us consider the example of an ideal gas. The change in its Helmholtz free energy, is given by $d\mathcal{A} = d\mathcal{U} - 
Td\mathcal{S} - \mathcal{S}dT$. Using the first and the second laws of thermodynamics we have $d\mathcal{U} = Td\mathcal{S} - pdV$. Therefore, $d\mathcal{A} = -pdV - \mathcal{S}dT$, which 
under isothermal conditions means $d\mathcal{A} = -pdV$. In this simple system, $p$ is the isotropic portion of the stress and $dV$ is related to the isotropic strain. Thus, we can get $p$ 
is we know change in Helmholtz free energy and volume.\footnote{Gibbs free energy gives the strain in terms of stress.}

Therefore, a 
first step toward getting an expression for stress is to find the Helmholtz free energy. Under isothermal conditions, the first law of thermodynamics is
\begin{equation}\label{fe:1}
d\mathcal{U}=Td\mathcal{S}+\dbar\mathcal{W},
\end{equation}
where $\mathcal{U}$ is the total free energy, $T$ is the absolute temperature, $\mathcal{S}$ is the total entropy and $\mathcal{W}$ is the work done on the system. First law of 
thermodynamics for polarizable and magnetizable media is 
\begin{eqnarray}
\delta\mathcal{U} &=& T\delta\mathcal{S} + \dbar\mathcal{W}_0 + \int\vec{E}\cdot d\vec{D}dV	\label{fe:2}	\\
\delta\mathcal{U} &=& T\delta\mathcal{S} +\dbar\mathcal{W}_0 + \int\vec{H}\cdot d\vec{B}dV	\label{fe:3},
\end{eqnarray}
where $\mathcal{W}_0$ is the mechanical work done on the system. If $U$, $W_0$ and $S$ are internal energy, mechanical work and entropy of the media \emph{per unit volume}, first law of 
thermodynamics for polarizable and magnetizable media is
\begin{eqnarray}
\delta{U} &=& T\delta{S} + \dbar{W_0} + \vec{E}\cdot\delta\vec{D}	\label{fe:4}	\\
\delta{U} &=& T\delta{S} + \dbar{W_0} + \vec{H}\cdot\delta\vec{B}	\label{fe:5}	
\end{eqnarray}
The mechanical work done on a material is $\delta W_0=-\pi_{ij}d \gamma_{ij}$ where $\dpi$ is the stress tensor and $\gamma_{ij}$ is the strain tensor in the medium\footnote{Equation (3.1) 
of \cite{ch2:ll7}}. Further, with this substitution, all quantities in equations (\ref{fe:4}) and (\ref{fe:5}) become exact differentials allowing us to replace $\text{ }\dbar$ with $d$. 
If $\breve{A}$ is the Helmholtz free energy per unit volume,
\begin{eqnarray}
d\breve{A} &=& -SdT-\pi_{ij}d \gamma_{ij} + \vec{E}\cdot d\vec{D}		\label{fe:6}	\\
d\breve{A} &=& -SdT-\pi_{ij}d \gamma_{ij} + \vec{H}\cdot d\vec{B}		\label{fe:7}
\end{eqnarray}
These relations give change in Helmholtz free energy in terms of change in $\vec{D}$ and $\vec{B}$. The $\vec{D}$ field's source is free charges alone while the $\vec{B}$ field's source is 
all currents. In an experiment, we can control the total charge and free currents. Therefore, it is convenient to express free energy in terms of $\vec{E}$, whose source is all charges -
free and bound, and $\vec{H}$, whose source is free currents. We therefore introduce associated Helmholtz free energy function ${A}$ for polarizable media as ${A} = \breve{A}-\vec{E}
\cdot\vec{D}$ and for magnetizable media as ${A} = \breve{A}-\vec{H}\cdot\vec{B}$. Equations (\ref{fe:6}) and (\ref{fe:7}) therefore become
\begin{eqnarray}
d{A} &=&-SdT - \pi_{ij}d \gamma_{ij} - \vec{D}\cdot d\vec{E}		\label{fe:8}	\\
d{A} &=&-SdT - \pi_{ij}d \gamma_{ij} - \vec{B}\cdot d\vec{H}		\label{fe:9}
\end{eqnarray}
If $\dtau$ is the deviatoric stress, $\dpi = p\udyad + \dtau$, where $p$ is the hydrostatic pressure. Therefore we have,
\begin{eqnarray}
d{A} &=&-SdT - p d\gamma_{ii} - \tau_{ij}d \gamma_{ij} - \vec{D}\cdot d\vec{E}		\label{fe:8a}	\\
d{A} &=&-SdT - p d\gamma_{ii} - \tau_{ij}d \gamma_{ij} - \vec{B}\cdot d\vec{H}		\label{fe:9a}
\end{eqnarray}
The quantity $d\gamma_{ii}$ is the dilatation\footnote{Ratio of change in volume to the original volume.} of the material during deformation. Therefore the thermodynamic potential ${A}$ of 
a polarizable (magnetizable) medium is thus, a function of $T$, $V$, $\gamma_{ij}$ and $\vec{E}$($\vec{H}$). Equivalently, it can be considered a function of $T$, $\rho$, $\gamma_{ij}$ 
and $\vec{E}$($\vec{H}$), where $\rho$ is the mass density of the medium. 

\section{Stress in a dielectric viscoelastic liquid in electric field}\label{sec:dielectric}
We will now calculate the stress tensor in a polarizable medium. We consider a small portion of the material and find out the work done by the portion in a deformation in presence of an
electric field. The portion is small enough to approximate the field to be uniform throughout its extent. We emphasize that through this assumption we are not ruling out non-uniform
fields but only insisting that the portion be small enough to ignore variations in it. Since a sufficiently small portion of a material can be considered to be plane, the volume element
under consideration can be assumed to be in form of a rectangular slab of height $h$. Let it be subjected to a virtual displacement $\vec{\xi}$ which need not be parallel to the normal 
$\un$ to the surface. The virtual work done by the medium per unit area in this deformation is $\xi_i \sigma_{ij} n_j$, where $\dsigma$ is the stress \emph{on} the portion. If 
$\dpi$ is the stress \emph{due to} the portion \emph{on} the medium, then $\pi_{ij} = -\sigma_{ji}$. Therefore, the virtual work done by the medium on the portion is  
$-\xi_i\pi_{ji}n_j$. Further, since both $\dsigma$ and $\dpi$ are symmetric, the virtual work can also be written as $-\xi_i\pi_{ij}n_j$. The change in Helmholtz free energy 
during the deformation is $\delta({h}{A})$ per unit surface area. If we assume the deformation to be isothermal,
\begin{equation}\label{dielectric:1}
\xi_i \pi_{ij} n_j = -\delta(h{A}) = -(\delta h){A} - h(\delta{A})
\end{equation}
Change in height of the slab is 
\begin{equation}\label{dielectric:2}
\delta h = \vec{\xi}\cdot\un=\xi_i n_j \delta_{ij}
\end{equation}
\begin{figure}[!ht]\label{dielectric:f1}
\centering
\includegraphics[height=60mm, width=100mm]{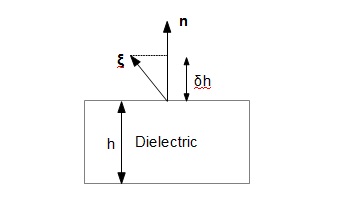}
\caption{A virtual displacement of a dielectric material}
\end{figure}
The geometry of the problem is described in figure 2. For an isothermal variation
\begin{equation}\label{dielectric:3}
\delta{A}=\left(\pdt{{A}}{\rho}\right)_{\gamma_{ij},E_i}\delta\rho+\left(\pdt{{A}}{ \gamma_{ij}}\right)_{\rho,E_i}\delta\gamma_{ij} +\left(\pdt{{A}}{E_i}\right)_{\rho,\gamma_{ij}}
\delta E_i
\end{equation}
We depart from the convention in thermodynamics, to indicate variables held constant as subscripts to partial derivatives, to make our equations appear neater. We shall also use the 
traditional notation for partial derivatives. We will now get expressions for each term on the right hand side of equation (\ref{dielectric:3}).
\begin{enumerate}
\item If $A_0$ is the Helmholtz free energy in absence of electric field, ${A} = A_0 - \smallint\vec{D}\cdot d\vec{E} = A_0 - \smallint\epsilon_{ij}E_i dE_j$, where $\epsilon_{ij}$ 
is the permittivity tensor. Permittivity is known to be a function of mass density of a material, the dependence being given by Clausius-Mossotti relation\cite{ch2:pp}. Electric field is 
\emph{usually} independent of mass density of the material. However, that is not so if the material has a pronounced density stratification like a fluid heated from above. If $V_t$ and 
$V_b$ are two elements of such a fluid, at the top and bottom respectively, both having identical volume then $V_t$ will have less number of dipoles than $V_b$. The electric field inside 
them, due to matter within their confines too will differ. We point out that although divergence of $\vec{E}$ depends only on the density of free charges, $\vec{E}$ itself is produced by 
all multipoles. Therefore,
\begin{equation}\label{dielectric:4}
\pdt{{A}}{\rho}=\pdt{A_0}{\rho}-\pdt{\epsilon_{ij}}{\rho}\int E_j dE_i-\epsilon_{ij}\frac{\partial}{\partial\rho}\left(\int E_j dE_i\right)
\end{equation}
The last term in equation (\ref{dielectric:4}) is absent if the material has a uniform temperature. It is not included in the prior works of Bobbio\cite{ch2:bobbio} and Landau and
Lifshitz\cite{ch2:ll8}. 

\item If the $x_3$ (or $z$) axis is assumed to be along the normal and the deformation is uniform, the displacement of a layer of the volume element can be described as 
\begin{equation}\label{dielectric:5}
\vec{s}=\frac{x_3\vec{\xi}}{h},
\end{equation}
where $x_3$ is the vertical distance from the lower surface. Since $\vec{\xi}$ is fixed,
\begin{equation}\label{dielectric:6}
\pdt{s_i}{x_j}=\frac{n_j\xi_i}{h},
\end{equation}
and
\begin{equation}\label{dielectric:7}
\delta\gamma_{ij} = \left(\pdt{s_j}{x_i}+\pdt{s_i}{x_j}\right)=\frac{1}{h}(\xi_j n_i + \xi_i n_j)
\end{equation}
Since the strain tensor is always symmetric\footnote{Proved in appendix to this paper.},
\begin{equation}\label{dielectric:8}
\pdt{A_0}{ \gamma_{ij}}\delta \gamma_{ij}=\frac{2\xi_i n_j}{h}\pdt{A_0}{\gamma_{ij}}
\end{equation}
The electric field does not depend on strain but permittivity does. This is because, deformation may change the anisotropy of the material, which determines its permittivity. 
Likewise, permittivity and strain tensors do not depend on electric field\footnote{Let us assume a linear material for the moment. We will relax it in section \ref{sec:ferro}}. Therefore they can be pulled out of the integral and 
\begin{equation}\label{dielectric:9}
\pdt{{A}}{ \gamma_{ij}}\delta  \gamma_{ij}=\frac{\xi_i n_j}{h}\left(2\pdt{A_0}{ \gamma_{ij}}-\pdt{\epsilon_{rs}}{ \gamma_{ij}}\int E_s dE_r\right)
\end{equation}

\item  From equation (\ref{fe:8}),
\begin{equation}\label{dielectric:10}
\left(\pdt{{A}}{E_i}\right) = -D_i
\end{equation}
Therefore, the last term is just $-\vec{D}\cdot d\vec{E}$.
\end{enumerate}

Using equations (\ref{dielectric:4}), (\ref{dielectric:9}) and (\ref{dielectric:10}) in (\ref{dielectric:3}), we get
\begin{eqnarray}
\delta{A} &=& \left(\pdt{A_0}{\rho}-\pdt{\epsilon_{ij}}{\rho}\int E_jdE_i-\epsilon_{ij}\frac{\partial}{\partial\rho}\left(\int E_j dE_i\right)\right)\delta\rho	+ \nonumber		\\
	& & \frac{\xi_i n_j}{h}\left(2\pdt{A_0}{\gamma_{ij}}-\pdt{\epsilon_{rs}}{ \gamma_{ij}}\int E_s dE_r\right)-\vec{D}\cdot\delta\vec{E}	\label{dielectric:11}
\end{eqnarray}
Substituting (\ref{dielectric:11}) and (\ref{dielectric:2}) in (\ref{dielectric:1}) we get
\begin{eqnarray}\label{dielectric:12}
-\xi_i \pi_{ij} n_j &=& \left\{A_0\xi_i n_j\delta_{ij} + h\delta\rho\pdt{A_0}{\rho} + 2\xi_i n_j\pdt{A_0}{\gamma_{ij}}\right\} - \nonumber	\\
 & & h\delta\rho\pdt{\epsilon_{ij}}{\rho}\int E_jdE_i - h\delta\rho\epsilon_{ij}\frac{\partial}{\partial\rho}\left(\int E_j dE_i\right) -	\nonumber \\
 & & \xi_i n_j \pdt{\epsilon_{rs}}{ \gamma_{ij}}\int E_s dE_r - \xi_i n_j\delta_{ij}\int\vec{D}\cdot d\vec{E} -\nonumber	\\
 & & h\vec{D}\cdot\delta\vec{E} 
\end{eqnarray}
We have gathered terms independent of electric field in the first curly bracket of equation (\ref{dielectric:12}), keeping the contribution of electric field to stress in the rest. We 
still have to find out the expressions for $\delta\rho$ and $\delta\vec{E}$. A change in density of a layer depends on the change it its height (or thickness), therefore, $\delta\rho = 
-(\delta h/h)\rho=-(\vec{\xi}\cdot\un/h)\rho$ or,
\begin{equation}\label{dielectric:12a}
h\delta{\rho} = -\rho(\vec{\xi}\cdot\un) = -\rho\xi_i n_j \delta_{ij}
\end{equation}
We will now estimate change in electric field due to deformation. Consider a volume element at a point $\vec{x}$. Let it undergo a deformation by $\vec{s}$. As a result, matter that used 
to be at $\vec{x}-\vec{s}$ now appears at $\vec{x}$. In a virtual homogeneous deformation, every volume element carries its potential as the material deforms. Therefore, the change in 
potential at $\vec{x}$ is 
$\delta\phi= \phi(\vec{x}-\vec{s})-\phi(\vec{x})=-\vec{s}\cdot\grad{\phi}=\vec{s}\cdot\vec{E}$. Since $\vec{s}=(x_3\vec{\xi})/h$ (see equation (\ref{dielectric:5})),
\begin{equation}\label{dielectric:13}
\delta\phi=\frac{x_3\vec{\xi}}{h}\cdot\vec{E}
\end{equation}
Since $\grad{x_3}=\un$, 
\begin{equation}\label{dielectric:14}
\delta\vec{E}=-\grad{(\delta\phi)}=-\frac{\un(\vec{E}\cdot\vec{\xi})}{h}
\end{equation}
We have used the assumption that the region is small enough to have almost uniform electric field and therefore it can be pulled out of the gradient operator.
Equation (\ref{dielectric:12}) therefore becomes
\begin{eqnarray}\label{dielectric:15}
-\xi_i \pi_{ij} n_j &=& \xi_i n_j\left\{A_0\delta_{ij} - \rho\delta_{ij}\pdt{A_0}{\rho} + 2\pdt{A_0}{ \gamma_{ij}}\right\} + \nonumber	\\
 & & \xi_i n_j\left\{\rho\delta_{ij}\pdt{\epsilon_{rs}}{\rho}\int E_sdE_r + \rho\delta_{ij}\epsilon_{rs}\frac{\partial}{\partial\rho}\int E_s dE_r\right\} - \nonumber	\\  
 & & \xi_i n_j\left\{\pdt{\epsilon_{rs}}{ \gamma_{ij}}\int E_sdE_r - E_i D_j + \delta_{ij}\int\vec{D}\cdot d\vec{E}	\right\} 
\end{eqnarray}
Stress in a polarized viscoelastic material at rest is therefore,
\begin{eqnarray}\label{dielectric:16a}
\pi_{ij} &=& - \left\{A_0\delta_{ij} - \rho\delta_{ij}\pdt{A_0}{\rho} + 2\pdt{A_0}{ \gamma_{ij}}\right\}  \nonumber	\\
 & & - \left\{\rho\delta_{ij}\pdt{\epsilon_{rs}}{\rho}\int E_sdE_r + \rho\delta_{ij}\epsilon_{rs}\frac{\partial}{\partial\rho}\int E_s dE_r\right\}  \nonumber	\\  
 & & + \left\{\pdt{\epsilon_{rs}}{ \gamma_{ij}}\int E_sdE_r - E_i D_j + \delta_{ij}\int\vec{D}\cdot d\vec{E}	\right\} 
\end{eqnarray}
We can simplify equation (\ref{dielectric:16a}) by writing the terms in the first curly bracket as familiar thermodynamic quantities. If $\mathcal{A}_0$ is the total Helmholtz free energy
of the substance in absence of electric field and $A_0$ is the Helmholtz free energy per unit volume then $\mathcal{A}_0=A_0 V$, where $V=M/\rho$ is the volume of the substance, $M$ the 
mass and $\rho$ the density. Maxwell relation for pressure in terms of total Helmholtz free energy is
\begin{equation}\label{dielectric:16b}
p=-\left(\pdt{\mathcal{A}_0}{V}\right)_T=-A_0+\rho\left(\pdt{A_0}{\rho}\right)_T
\end{equation}
Similarly, the dependence of $A_0$ on strain tensor $ \gamma_{ij}$ can be written as $(G/2)( \gamma_{ij}-(1/3) \gamma_{kk}\delta_{ij})^2$\cite{ch2:ll7}, where we have retained only the 
deviatoric of the strain tensor because the isotropic part is already accounted in hydrostatic pressure of equation (\ref{dielectric:16a})\footnote{We are ignoring the possibility of an
elastic deformation either caused by or causing a temperature gradient.}. The constant $G$ is the shear modulus of the 
substance. Therefore,
\begin{equation}\label{dielectric:16c}
2\pdt{A_0}{ \gamma_{ij}} = G\left(\gamma_{ij}-\frac{1}{3} \gamma_{kk}\delta_{ij}\right)
\end{equation}
Equation (\ref{dielectric:16a}) can therefore be written as
\begin{eqnarray}\label{dielectric:16}
\pi_{ij} &= & \left\{p\delta_{ij} - G\left( \gamma_{ij}-\frac{1}{3} \gamma_{kk}\delta_{ij}\right)\right\} -	\nonumber \\
 & & \left\{\rho\delta_{ij}\pdt{\epsilon_{rs}}{\rho}\int E_sdE_r + \rho\delta_{ij}\epsilon_{rs}\frac{\partial}{\partial\rho}\int E_s dE_r\right\} + \nonumber	\\  
 & & \left\{\pdt{\epsilon_{rs}}{ \gamma_{ij}}\int E_sdE_r - E_i D_j + \delta_{ij}\int\vec{D}\cdot d\vec{E}	\right\} 
\end{eqnarray}
We will now look at some special cases of (\ref{dielectric:16}),
\begin{enumerate}
\item If there is no matter, terms with pressure, density and strain tensor will not be present. Further $\epsilon_{rs}=\epsilon_0\delta_{rs}$ and equation (\ref{dielectric:16}) becomes 
the Maxwell stress tensor for electric field in vacuum.
\begin{equation}\label{dielectric:17a}
\pi_{ij} = \epsilon_0\left(\frac{E^2}{2}\delta_{ij} - E_iE_j\right)
\end{equation}
We emphasize that the general expression for stress in a material exposed to static electric field reduces to Maxwell stress tensor only when we ignore all material properties. 
\item If there is no electric field, all terms in the second and third curly bracket of (\ref{dielectric:16}) vanish. Further, if the medium is a fluid without elastic properties, $A_0$, 
will not depend on $ \gamma_{ij}$ and the stress will be
\begin{equation}\label{dielectric:17}
\pi_{ij} = \left(\rho\pdt{A_0}{\rho} - A_0\right)\delta_{ij} = p\delta_{ij}
\end{equation}
Thus the stress in a fluid without elastic properties is purely hydrostatic. We do not see viscous terms in (\ref{dielectric:17}) because viscosity is a dissipative effect while 
$\pi_{ij}$ is obtained from Helmholtz free energy which has information only about energy than can be extracted as work. 
\item If the material were a solid and if there are no electric fields as well, the stress is
\begin{equation}\label{dielectric:18}
\pi_{ij} = p\delta_{ij} + G\left(\frac{1}{3} \gamma_{kk}\delta_{ij} -  \gamma_{ij}\right)
\end{equation}
It is customary to write the first term of equation (\ref{dielectric:18}) in terms of $K$, the bulk modulus as
\begin{equation}\label{dielectric:18a}
\pi_{ij} = -K \gamma_{kk}\delta_{ij}+ G\left(\frac{1}{3} \gamma_{kk}\delta_{ij} -  \gamma_{ij}\right)
\end{equation}
\item For a fluid dielectric with isotropic permittivity tensor, $A_0$ is independent of $ \gamma_{ij}$ and $\epsilon_{rs}=\epsilon\delta_{rs}$. If the fluid has a uniform density, equation 
(\ref{dielectric:16}) then becomes
\begin{equation}\label{dielectric:19}
\pi_{ij}= p\delta_{ij} + \left(\frac{\epsilon}{2}E^2 - \rho\frac{E^2}{2}\pdt{\epsilon}{\rho}\right) \delta_{ij} - \epsilon E_i E_j
\end{equation}
This expression matches the one obtained in~\cite{ch2:ll8}, after converting to Gaussian units, and after accounting for the difference in the interpretation of stress tensor. Landau and
Lifshitz's stress tensor is $\sigma_{ij} = -\pi_{ji}$.
\item For a fluid dielectric with isotropic permittivity tensor and in which the electric field depends on density equation (\ref{dielectric:16}) then becomes
\begin{equation}\label{dielectric:19a}
\pi_{ij}= p\delta_{ij} + \left(\frac{\epsilon}{2}E^2 - \rho\frac{E^2}{2}\pdt{\epsilon}{\rho} - \frac{\rho\epsilon}{2}\pdt{(E^2)}{\rho}\right) \delta_{ij} - \epsilon E_i E_j
\end{equation}
\item For a solid dielectric we can assume that $A_0$, $\vec{E}$ and $\epsilon_{rs}$ are independent of $\rho$. Equation (\ref{dielectric:16}) now becomes
\begin{equation}\label{dielectric:20}
\pi_{ij} = -A_0\delta_{ij} - \pdt{A_0}{ \gamma_{ij}} + \left\{\pdt{\epsilon_{rs}}{ \gamma_{ij}}\int E_s dE_r + \epsilon_{rs}\int E_r dE_s\delta_{ij} - D_iE_j\right\}
\end{equation}
If the solid is isotropic and remains to be so after application of electric field, $\epsilon_{rs}=\epsilon\delta_{rs}$ and (\ref{dielectric:20}) simplifies to
\begin{equation}\label{dielectric:21}
\pi_{ij} = \pi^{0}_{ij} + \epsilon\left(\frac{E^2}{2}\delta_{ij} - E_i E_j\right) + \frac{E^2}{2}\pdt{\epsilon}{ \gamma_{ij}},
\end{equation}
where
\begin{equation}\label{dielectric:22}
\pi^{0}_{ij} = -K  \gamma_{kk}\delta_{ij} + 2\mu \left(\frac{1}{3} \gamma_{kk}\delta_{ij} -  \gamma_{ij}\right)
\end{equation}
is the part of stress tensor that exists even in absence of electric field. This expression matches with the one in~\cite{ch2:ll8} if one converts it to Gaussian units, assumes the 
constitutive relation $\vec{D}=\epsilon( \gamma_{ij})\vec{E}$ and takes into account that their stress tensor is $\sigma_{ij} = -\pi_{ji}$.
\item For a viscoelastic liquid that is also a linear dielectric with uniform density,
\begin{equation}\label{dielectrc:23}
\pi_{ij} = p\delta_{ij} - 2\pdt{A_0}{ \gamma_{ij}}+\frac{E^2}{2}\left(\pdt{\epsilon}{ \gamma_{ij}}-\rho\pdt{\epsilon}{\rho}\right)+\epsilon\left(\frac{E^2}{2}\delta_{ij} - E_iE_j\right)
\end{equation}
\end{enumerate}

\section{Stress in a magnetic viscoelastic liquid in magnetic field}\label{sec:magnetic}
In order to calculate stress in a magnetic fluid, we continue to use the physical set up used in section (\ref{sec:dielectric}) of a small slab of viscoelastic liquid subjected to magnetic
field. If there are no conduction and displacement currents, Ampere's law becomes $\curl{\vec{H}}=0$, making the $\vec{H}$ field conservative. It can then be treated like the electrostatic
field of section (\ref{sec:dielectric}). In order to extend the analysis of section (\ref{sec:dielectric}) to magnetic fluids, we need an additional assumption of magnetic permeability
being independent of $\vec{H}$. Although the first assumption, of no conduction and displacement currents, is valid in the case of ferro-viscoelastic fluids, the second assumption of 
field-independent permeability is not. Therefore, this analysis is valid only for the single-valued, linear section of the $\vec{B}$ versus $\vec{H}$ curve of ferro-viscoelastic liquid,
giving
\begin{eqnarray}\label{magnetic:1}
\pi_{ij} &= & \left\{p\delta_{ij} - G\left( \gamma_{ij} - \frac{1}{3} \gamma_{kk}\delta_{ij}\right)\right\} -	\nonumber \\
 & & \left\{\rho\delta_{ij}\pdt{\mu_{rs}}{\rho}\int H_s dH_r + \pdt{\mu_{rs}}{ \gamma_{ij}}\int H_s dH_r\right\} + \nonumber	\\  
 & & \left\{\delta_{ij}\int\vec{B}\cdot d\vec{H} - H_i B_j	\right\},
\end{eqnarray}
where $\mu_{rs}$ is the magnetic permeability tensor. We have omitted the term accounting for dependence of $\vec{H}$ on mass density $\rho$ because we are not aware of a situation where
it may happen.

\section{Stress in permanently magnetized or polarized media}\label{sec:ferro}
The expressions derived in sections (\ref{sec:dielectric}) and (\ref{sec:magnetic}) are valid only if permittivity and permeability are independent of electric and magnetic fields 
respectively. Ferro-fluids are colloids of permanently magnetized particles. As the applied magnetic field increases from zero, an increasing number of sub-domain magnetic particles align
themselves parallel to the field opposing the random thermal motion leading to a magnetization that increased in a non-linear manner. The magnetic susceptibility and therefore permeability
depend on the field. It cannot be pulled out of the integral sign. Equation (\ref{magnetic:1}) should be written as
\begin{eqnarray}\label{ferro:1}
\pi_{ij} &= & \left\{p\delta_{ij} - G\left( \gamma_{ij} - \frac{1}{3} \gamma_{kk}\delta_{ij}\right)\right\} -	\nonumber \\
 & & \left\{\delta_{ij}\int\rho\pdt{\mu_{rs}}{\rho} H_s dH_r - \int \pdt{\mu_{rs}}{ \gamma_{ij}} H_s dH_r\right\} + \nonumber	\\  
 & & \left\{\delta_{ij}\int\mu_{rs}H_s dH_r - H_i B_j	\right\}
\end{eqnarray}
If the elastic effects are negligible, equation (\ref{ferro:1}) reduces to
\begin{equation}\label{ferro:2}
\pi_{ij} = p\delta_{ij} + \delta_{ij}\left\{\int\left(\mu_{rs}-\rho\pdt{\mu_{rs}}{\rho}\right)H_rdH_s\right\} - H_i B_j
\end{equation}
If $\mu_{rs}=\mu\delta_{rs}$, as is normally for ferro-fluids~\cite{ch2:rer},
\begin{equation}\label{ferro:3}
\pi_{ij} = p\delta_{ij} + \delta_{ij}\left\{\int\mu H_rdH_r - \int\rho\pdt{\mu}{\rho}H_rdH_r\right\} - H_i B_j
\end{equation}
Since the applied magnetic field is independent of density,
\begin{equation}\label{ferro:4}
\rho\pdt{\mu}{\rho}H_r=\rho\pdt{(\mu H_r)}{\rho}=\rho\pdt{B_r}{\rho}=\mu_0\rho\pdt{M_r}{\rho},
\end{equation}
where to get the last equation we have used the relation $\vec{B}=\mu_0(\vec{H}+\vec{M})$ and the fact that $\vec{H}$ does not depend on $\rho$. If $v$ is the specific volume, that is 
$v=1/\rho$, equation (\ref{ferro:5}) can be written as
\begin{equation}\label{ferro:5}
\rho\pdt{\mu}{\rho}H_r=-\mu_0 v\pdt{M_r}{v}
\end{equation}
Further, $\mu\vec{H}=\vec{B}$ and $\vec{B}=\mu_0(\vec{H}+\vec{M})$ imply,
\begin{equation}\label{ferro:6}
\mu H_r = \mu_0 H_r + \mu_0 M_r
\end{equation}
Using equations (\ref{ferro:5}) and (\ref{ferro:6}) in equation (\ref{ferro:3}), we get
\begin{equation}\label{ferro:7}
\pi_{ij} = \left\{p + \frac{\mu_0}{2}H^2 + \mu_0\int\left[\pdt{(vM_r)}{v}\right]dH_r\right\}\delta_{ij} - H_i B_j
\end{equation}
This is same as Rosensweig's~\cite{ch2:rer} equation (4.28) except that he calculates $\dsigma$, which is related to our stress tensor as $\sigma_{ij} = -\pi_{ji}$. We do not 
know of electric analogues of ferro fluids (electro-rheological fluids are analogues of magneto-rheological fluids, not ferro fluids).  However, there are permanently polarized solids, 
called ferro-electrics. For such materials, the stress is
\begin{equation}\label{ferro:x}
\pi_{ij} = -A_0\delta_{ij} - 2\pdt{A_0}{ \gamma_{ij}} + \left\{\int\pdt{\epsilon_{rs}}{ \gamma_{ij}}E_sdE_r + \int\epsilon_{rs}E_r dE_s\delta_{ij} - E_i D_j\right\}
\end{equation}

\section{Ponderomotive forces}\label{sec:ponder}
The old term "ponderable media" means media that have weight. Ponderomotive force is the one that cause motion or deformation in a ponderable medium. In contemporary terms, it is the 
density of body force in a material. It is related to the stress tensor $\dpi$ as
\begin{equation}\label{ponder:1}
f_i = -\pi_{ij,j}
\end{equation}
We mention a few familiar special cases of this equation for fluids of various kinds.
\begin{enumerate}
\item For incompressible, Newtonian fluids the stress tensor is given by (\ref{dielectric:17}) and the force density is
\begin{equation}\label{ponder:2}
\vec{f}=-\grad{p}.
\end{equation}
Note that the force density does not include the dissipative component due to viscosity.
\item For an incompressible, Newtonian, dielectric fluid in presence of static electric field, assuming that the electric field inside the fluid is independent of density, the stress 
tensor is given by equation (\ref{dielectric:19}) and the ponderomotive force is
\begin{equation}\label{ponder:3}
\vec{f}=-\grad{p}+\rho_f\vec{E}-\frac{\epsilon_0 E^2}{2}\grad{\kappa}+\frac{\epsilon_0}{2}\grad{\left(E^2\rho\frac{d\kappa}{d\rho}\right)},
\end{equation}
where $\rho_f$ is the density of free charges in the fluid and $\kappa$ is its relative permittivity. In deriving equation (\ref{ponder:3}) we used Gauss' law $\dive{\vec{D}}=\rho_f$ and 
the fact that we are dealing with an electrostatic field ($\curl{\vec{E}}=0$), for which $\grad({E^2/2})=\vec{E}\cdot\grad{\vec{E}}$. In an ideal, dielectric fluid $\rho_f=0$ and
\begin{equation}\label{ponder:4}
\vec{f}=-\grad{p}-\frac{\epsilon_0 E^2}{2}\grad{\kappa}+\frac{\epsilon_0}{2}\grad{\left(E^2\rho\frac{d\kappa}{d\rho}\right)},
\end{equation}
The relative permittivity is a function of temperature and the term $\grad{\kappa}$ is significant in a single-phase fluid only if there is a temperature gradient. The third term in 
equation (\ref{ponder:4}) is called the electro-striction term and it is present only when the electric field or $\rho(d\kappa/d\rho)$ or both are non-uniform. The derivative of the 
relative permittivity with respect to mass density is calculated using the Clausius-Mossotti relation~\cite{ch2:asj},~\cite{ch2:pp}. 
\item Continuing with the same fluid as above but now having a situation in which the electric field is a function of mass density $\rho$, we have an additional term in equation 
(\ref{ponder:4}) given by
\begin{equation}\label{ponder:4a}
\vec{f} = -\grad{p}-\frac{\epsilon_0 E^2}{2}\grad{\kappa}+\frac{\epsilon_0}{2}\grad{\left(\rho\frac{d}{d\rho}\left(\kappa E^2\right)\right)}
\end{equation}
We come across such a situation when there is a strong temperature gradient in the fluid resulting in a gradient of dielectric constant $\kappa$. Since the electric field depends on 
$\kappa$ and $\kappa$ depends on mass density through the Clausius-Mossotti relation, the electric field is a function of mass density and we have to consider this additional term. We
hasten to add that it not necessary for there to be a temperature gradient to have such a situation, a gradient of electric permittivity suffices to give rise to such a situation.
\item The derivation for force density in an incompressible, Newtonian, diamagnetic or paramagnetic fluid in presence of a static magnetic field is similar except that we use the Maxwell's
equations $\dive{\vec{B}}=0$ and $\curl{\vec{H}}=\vec{J}$. We also assume the auxiliary magnetic field, $\vec{H}$, inside the fluid is independent of density. We get
\begin{equation}\label{ponder:5}
\vec{f}=-\grad{p}+\vec{J}\vp\vec{B}-\frac{\mu_0}{2}H^2\grad{\kappa_m}+\frac{\mu_0}{2}\grad{\left(H^2\rho\frac{d\kappa_m}{d\rho}\right)}.
\end{equation}
The term $\vec{J}\vp\vec{B}$ is the Lorentz force term and it is zero if the fluid is not conducting. $\kappa_m$ is the relative permeability of the fluid. The fourth term in equation
(\ref{ponder:5}) is called the magneto-striction force. It is present only if the magnetic field or $\rho(d\kappa_m/d\rho)$ or both are non-uniform. The derivative of relative permeability 
with respect to mass density is calculated using the magnetic analog of the Clausius-Mossotti relation~\cite{ch2:pp}.
\item Several forms of body force density, all equivalent to each other, can be derived for ferro-fluids from equations (\ref{ferro:7}) and (\ref{ponder:1}). We refer to 
~\cite{ch2:rer} for more details.
\item\label{ponder:i5} If the material is dielectric and viscoelastic, we assume that the permittivity depends on the strain. Even though the material was isotropic before applying 
electric field, it may turn anisotropic as its molecules get polarized and align with the field. The scalar permittivity is then replaced with a second order permittivity tensor 
$\epsilon_{rs}$. Following Landau and Lifshitz's treatment of solid dielectrics~\cite{ch2:ll8}, we assume that the permittivity tensor is a linear function of the strain tensor and write 
it as
\begin{equation}
\epsilon_{rs} = \epsilon_0\delta_{rs} + b_1  \gamma_{rs} + b_2  \gamma_{tt}\delta_{rs},
\end{equation}
where $b_1$ and $b_2$ are constants indicating rate of change of permittivity with strain. We call them $b_1$ and $b_2$ to differentiate them from $a_1$ and $a_2$ used to describe behavior
of solid dielectric~\cite{ch2:ll8}. If we assume the material to be incompressible, $\gamma_{tt}=0$ and
\begin{equation}\label{ponder:7}
\pdt{\epsilon_{rs}}{ \gamma_{ij}}=0+b_1\delta_{ir}\delta_{js}
\end{equation}
For an incompressible material, equation (\ref{dielectric:16c}) becomes,
\begin{equation}\label{ponder:8}
\pdt{A_0}{ \gamma_{ij}} = \frac{G}{2}
\end{equation}
Since $b_1$ and shear modulus $G$ are constants, they do not survive in the expression for $\vec{f}$. The expression for ponderomotive force in a dielectric, viscoelastic fluid is same as 
that for a dielectric, Newtonian fluid.
\item\label{ponder:i6} The same conclusion follows for a viscoelastic fluid subjected to a magnetic field if we assume that $\mu_{rs} = \mu_0\delta_{rs} + c_1\gamma_{rs} + c_2 
\gamma_{tt}\delta_{rs}$, $c_1$ and $c_2$ being constants, when a fluid is magnetized.
\end{enumerate}

\section{Extension to time-varying fields}
Time-varying electric fields can penetrate conductors up to a few skin depths that depends on the frequency of the fields and physical parameters of the material like its ohmic 
conductivity or magnetic permeability. The general problem of response of materials to time-varying fields is quite complicated. However, the results in this paper can be applied for 
slowly varying fields, that is, the ones that do not significantly radiate. For such fields, the time varying terms of Maxwell equations can be ignored. Whether a time varying field can
be considered quasi-static or not depends on the linear dimension of the materials involved. If $\omega$ is the angular frequency of the fields, the wavelength of corresponding 
electromagnetic wave is $\lambda=(2\pi c)/\omega$, $c$ being the velocity of light in vacuum. If the linear dimension $L$ of the materials is much lesser than $\lambda$, for any element 
$d\vec{x}$ of the path of current, there is another within $L$ that carries same current in the opposite direction, effectively canceling the effect of current. For power line 
frequencies, the value of $L$ is a few hundred miles and even for low frequency radio waves, with $\omega=10^6$ Hz, $L$ is of the order of $30$ m. Thus the \emph{slowly-varying fields}
or \emph{quasi-static} approximation~\cite{ch2:rmc} is valid for frequencies up to that of radio waves and our results can be applied under those conditions.

\section{Appendix}
\begin{enumerate}
\item Proof of equation (\ref{dielectric:8}).
\begin{equation}\label{ch2app:e1}
\pdt{A_0}{\gamma_{ij}}\delta\gamma_{ij} = \frac{1}{h}\left(\pdt{A_0}{\gamma_{ij}}\xi_j n_i + \pdt{A_0}{\gamma_{ij}}\xi_i n_j\right)
\end{equation}
Interchange the indices in the first term,
\begin{equation}\label{ch2app:e2}
\pdt{A_0}{\gamma_{ij}}\delta\gamma_{ij} = \frac{1}{h}\left(\pdt{A_0}{\gamma_{ji}}\xi_i n_j + \pdt{A_0}{\gamma_{ij}}\xi_i n_j\right)
\end{equation}
Since $\gamma_{ij}$ is a symmetric tensor,
\begin{equation}\label{ch2app:e3}
\pdt{A_0}{\gamma_{ij}}\delta\gamma_{ij} = \frac{1}{h}\left(\pdt{A_0}{\gamma_{ij}}\xi_i n_j + \pdt{A_0}{\gamma_{ij}}\xi_i n_j\right) = \frac{2}{h}\pdt{A_0}{\gamma_{ij}}\xi_i n_j
\end{equation}
\end{enumerate}

\bibliographystyle{amsplain}

\end{document}